\documentclass[useAMS,usenatbib]{mnras}
\usepackage{graphicx}
\usepackage{float}
\usepackage{color}
\usepackage{upgreek}
\usepackage{comment}
\usepackage{multirow}
\usepackage{subfigure}

\def\mnras{Monthly Notices of the Royal Astronomical Society}

\title[A pilot survey for transients and variables with ASKAP]{A pilot survey for transients and variables with the Australian Square Kilometre Array Pathfinder}

\author[S. Bhandari et al.]{S. Bhandari$^{1,2,3}$\thanks{E-mail: shivani.bhandari@csiro.au}, 
K.W. Bannister$^{1}$, 
T. Murphy$^{4,3}$, 
M. Bell$^{14,1,3}$, 
W. Raja$^{1}$,
J. Marvil$^{1}$,\and
P. J. Hancock$^{6,3}$, 
M. Whiting$^{1}$,
C.M. Flynn$^{2,3}$,
J. D. Collier$^{1,8}$,
D. L. Kaplan$^{7}$, \and
J. R. Allison$^{4,13}$,
C. Anderson$^{1,9}$,
I. Heywood $^{11,12}$,
A. Hotan$^{1}$,
R. Hunstead$^{4}$,\and
K. Lee-Waddell$^{1}$, 
J. P. Madrid$^{1}$,
D. McConnell$^{1}$,
A. Popping$^{10,3}$, 
J. Rhee$^{10,3}$,\and
E. Sadler$^{3,4}$,
M. A. Voronkov$^{1}$ \\
\\
$^{1}$ CSIRO Astronomy and Space Science, PO Box 76, Epping, NSW 1710, Australia \\
$^{2}$ Centre for Astrophysics and Supercomputing, Swinburne University of Technology, Mail H30, PO Box 218, VIC 3122, Australia\\
$^{3}$ ARC Centre of Excellence for All-sky Astrophysics (CAASTRO)\\
$^{4}$ Sydney Institute for Astronomy, School of Physics, The University of Sydney, NSW 2006, Australia \\
$^{5}$ School of Mathematical and Physical Sciences, University of Technology, Sydney, NSW 2007, Australia  \\ 
$^{6}$ International Centre for Radio Astronomy Research, Curtin University, Bentley, WA 6102, Australia \\
$^{7}$ Department of Physics, University of Wisconsin--Milwaukee, Milwaukee, WI 53201, USA \\ 
$^{8}$ Western Sydney University, Locked Bag 1797, Penrith, NSW 2751, Australia. \\
$^{9}$ CSIRO Astronomy and Space Science, Perth, Australia \\
$^{10}$ International Centre for Radio Astronomy Research (ICRAR), UWA, 35 Stirling Highway, Crawley WA 6009, Australia. \\
$^{11}$ Astrophysics, Department of Physics, University of Oxford, Keble Road, Oxford OX1 3RH, UK\\
$^{12}$ Department of Physics and Electronics, Rhodes University, PO Box 94, Grahamstown, 6140, South Africa\\
$^{13}$ ARC Centre of Excellence for All-sky Astrophysics in 3 Dimensions (ASTRO 3D) \\
$^{14}$ University of Technology Sydney, 15 Broadway, Ultimo NSW 2007, Australia \\
}
\begin{document}

\date{Accepted . Received ; in original form }

\pagerange{\pageref{firstpage}--\pageref{lastpage}} \pubyear{2014}

\maketitle
\label{firstpage}

\begin{abstract}
We present a pilot search for variable and transient sources at 1.4~GHz with the Australian Square Kilometre Array Pathfinder (ASKAP). The search was performed in a 30~deg$^{2}$ area centred on the NGC 7232 galaxy group over 8 epochs and observed with a near-daily cadence. The search yielded nine potential variable sources, rejecting the null hypothesis that the flux densities of these sources do not change with 99.9\% confidence. These nine sources displayed flux density variations with modulation indices $m\geq0.1$ above our flux density limit of $\sim$1.5~mJy. They are identified to be compact AGN/quasars or galaxies hosting an AGN, whose variability is consistent with refractive interstellar scintillation. We also detect a highly variable source with modulation index $m> 0.5$ over a time interval of a decade between the Sydney University Molonglo Sky Survey (SUMSS) and our latest ASKAP observations. We find the source to be consistent with the properties of long-term variability of a quasar. No transients were detected on timescales of days and we place an upper limit $\rho_{t}<$~0.01~deg$^{-2}$ with 95$\%$ confidence for non-detections on near-daily timescales. The future VAST-Wide survey with 36-ASKAP dishes will probe the transient phase space with similar cadence to our pilot survey, but better sensitivity, and will detect and monitor rarer brighter events.
\end{abstract}


\section{Introduction}

In the past decade the variability of the radio sky has been investigated through a range of blind and targeted surveys.\footnote{\url{http://www.tauceti.caltech.edu/kunal/radio-transient-surveys/index.html}}
\addtocounter{footnote}{-1}
\begin{figure*}
\includegraphics[scale=0.6]{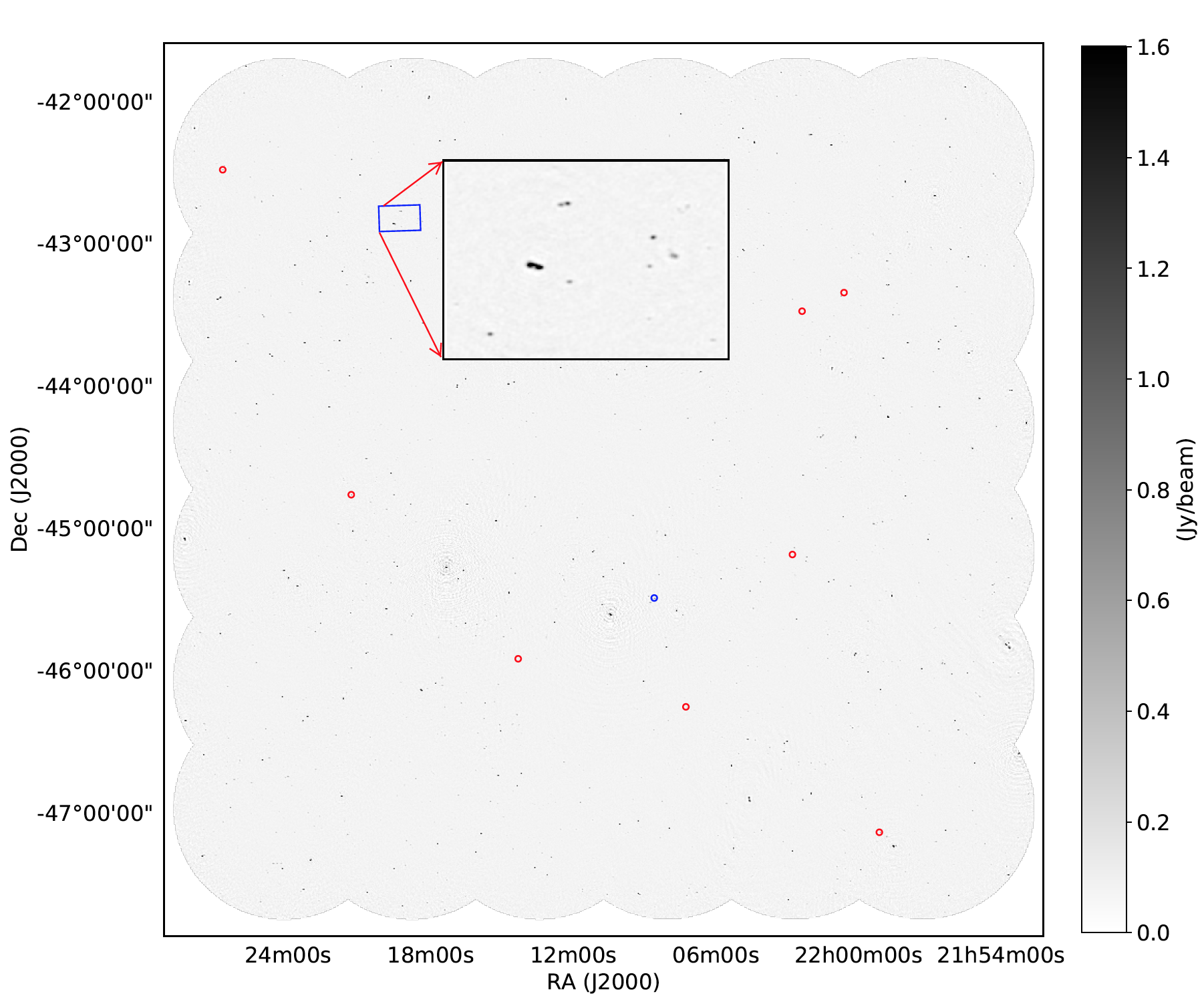}
\caption{A median image obtained by stacking seven
individual ASKAP epochs. Seven of the eight mosaicked images had the same image shape and therefore were stacked together. The image covers an area of 30~deg$^{2}$ at 1.4~GHz with an rms sensitivity of $\sim170~\upmu$Jy/beam. A total of 3817 sources were detected above a 5$\sigma$ detection limit. The red circles mark the sources detected to be potentially variable and the blue circle marks the highly variable source detected in our analysis.}
\label{figure:image}
\end{figure*}
These investigations have shown that the radio sky is relatively quiet for the sensitivities and timescales that have been explored so far, excluding the newly discovered class of millisecond transients called Fast radio bursts (FRBs). For example, \citet{Carilli2003} used the Karl G. Jansky Very Large Array (VLA) at 1.4~GHz to search for variable and transient sources in the Lockman hole. Their work showed that only 2\% of radio sources are highly ($>$50\%) variable above a peak flux density limit of 0.1~mJy on 19-day and 17-month timescales. 
\citet{bell} surveyed the $\sim$0.3 deg$^{2}$ area of the sky in the Chandra Deep Field South at 5.5~GHz and found 3$\%$ of the total sources to be variable. No transients were found, placing an upper limit of $<$ 7.5~deg$^{-2}$(95$\%$ confidence) above a detection threshold of 68.8~$\upmu$Jy. \citet{mooley} explored the variability at 3~GHz in a 50 deg$^{2}$ area of the sky in SDSS STRIPE 82 at a flux density limit of $\sim$0.5~mJy on timescales of weeks, months and years. Only 3.8\% of their sample had fractional variability more than 30$\%$ on timescales $<$~1.5 years. The fraction of point sources varying on week, month, and 1.5~year timescales were 1.0$\%$, 0.8$\%$ and 2.6$\%$ respectively. \citet{Hancock2016} measured the surface density of variable radio sources to be $\rho$ = 0.98~deg$^{-2}$ on time-scales of 6 months to 8 years in the Phoenix deep field above a flux density limit of 1~mJy at 1.4~GHz.

\begin{table*}
\centering
\caption{Summary of the 8 epochs, pointing B observations used in our analysis. Column 1 through 6 shows epoch number, scheduling block (SB) ID for each observation, date of observation, time spent on the target field, rms noise level for each image and FWHM of the synthesized beam. The observations were performed at the centre frequency of 1.4~GHz with a bandwidth of 48~MHz centred on RA:~22:08:03 and DEC:~$-$44:22:55 (J2000), covering 30~deg$^{2}$ on the sky.}
\label{table:Measurement1}
\begin{tabular}{|c|c|c|c|c|c|c|c|}
\hline 
Epoch & SB ID & Date  & T$_{\rm obs}$  & rms & Beam   \\
& & & (hrs) & ($\upmu$Jy/beam) & B$_{\rm max}$ $\times$ B$_{\rm min}$, PA \\
\hline
\hline
0 & SUMSS & 2003 Aug 06 & 12.0 & 1000 & \phantom{------}45$\arcsec~\rm{cosec}\delta \times 45\arcsec$, 0$^{\circ}$ \\
1 & 2238 &2016 Oct 07 & 10.4  &300  & 17$\arcsec \times$ 13$\arcsec$,\phantom{---}85$^{\circ}$ \\
2 & 2247 &2016 Oct 08 & 11.4  &320  & 17$\arcsec \times$ 13$\arcsec$,\phantom{---}85$^{\circ}$ \\
3 & 2253 &2016 Oct 09 & 12.1  &270  &18$\arcsec \times$ 12$\arcsec$,\phantom{---}87$^{\circ}$  \\
4 & 2264 & 2016 Oct 10 & 11.0  &360  &22$\arcsec \times$ 13$\arcsec,-$75$^{\circ}$  \\
5 & 2280 & 2016 Oct 12 & 7.3  &430  & 18$\arcsec \times$ 12$\arcsec$,\phantom{---}84$^{\circ}$   \\
6 & 2299 & 2016 Oct 14 & 10.4 &300  & 17$\arcsec \times$ 12$\arcsec$,\phantom{---}88$^{\circ}$   \\
7 & 2329 & 2016 Oct 17 & 12.3 &300  &18$\arcsec \times$ 12$\arcsec$,\phantom{---}87$^{\circ}$   \\
8 & 2347 & 2016 Oct 19 & 12.0 &300  & 18$\arcsec \times$ 12$\arcsec$,\phantom{---}86$^{\circ}$   \\
\hline
\end{tabular} 
\end{table*}
The surveys noted above do not give regular sampling of radio light curves
\citep[e.g.][]{Bannister2011}. In contrast, survey telescopes such as the Australian Square Kilometre Array Pathfinder \citep[ASKAP;][]{Johnston2007} will allow us to conduct consistently sampled surveys with wide fields of view (30~deg$^{2}$) and $\upmu$Jy sensitivity \citep{VAST}. Early exploration of this capability was performed by \citet{Hobbs2016} with the Boolardy Engineering Test Array \citep[BETA;][]{Hotan2014}. This resulted in no transient detection in 2~min snapshots in the field around pulsar PSR J1107$-$5907 above a flux density limit of 0.2~Jy. \citet{Heywood2016} also conducted a search for transients with ASKAP over 150~deg$^{2}$ with three epochs spanning a week and reported the detection of a significantly variable candidate quasar. 

In this paper we present the results of a pilot survey to search for variable and transient sources, using data obtained in the ASKAP Early Science program. In \S\ref{data reduction}, we describe observations and data processing using the ASKAPsoft\footnote{\url{http://www.atnf.csiro.au/computing/software/askapsoft/sdp/docs/current/index.html}} imaging pipeline. In \S\ref{analysis}, we present the image analysis. In \S\ref{results}, we describe our first variability results and upper limits on the transient source surface densities. In \S\ref{discussions} we discuss and summarize the implications of our results.

\begin{figure*}
\includegraphics[scale=0.4]{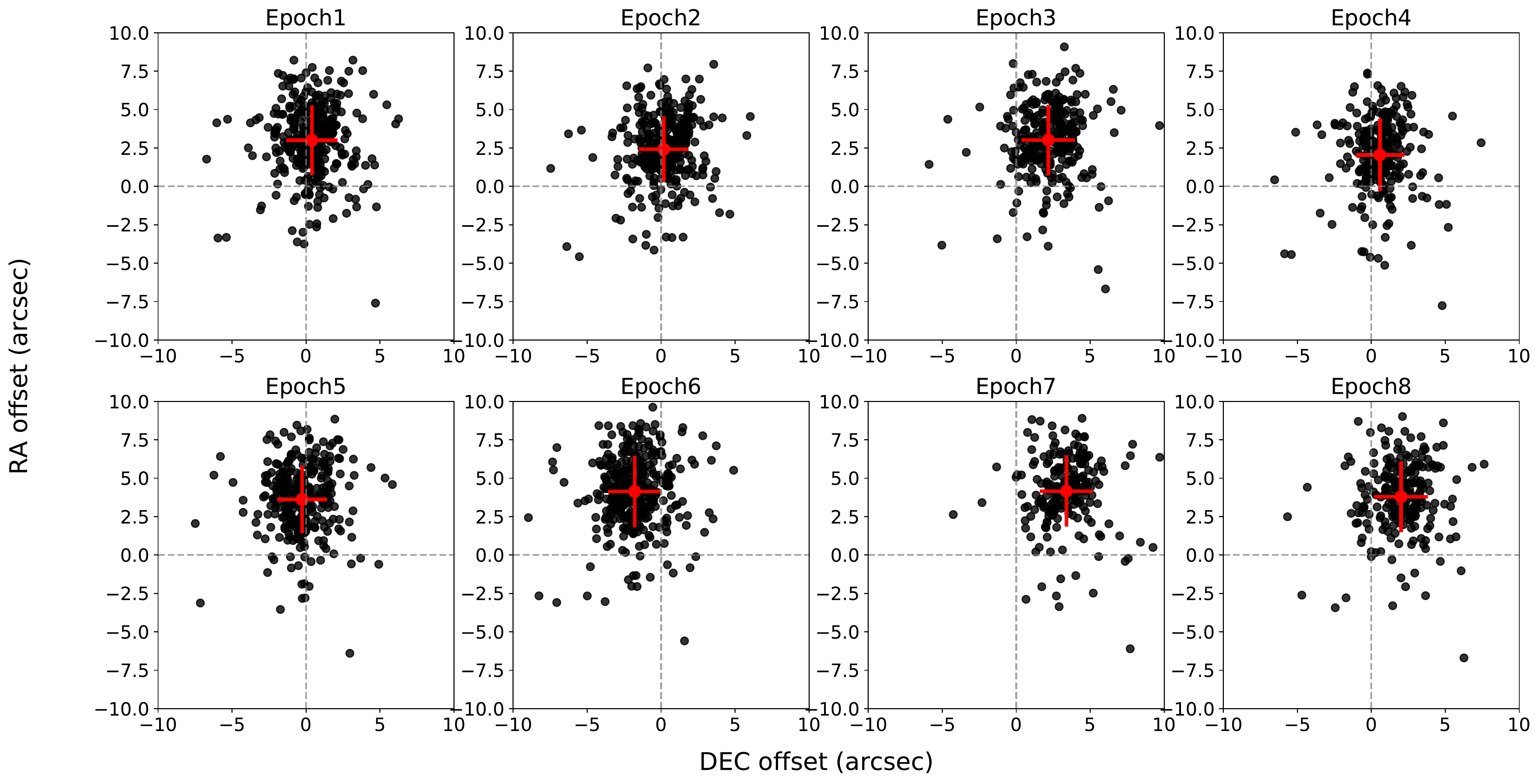}
\caption{Positional offsets of sources detected in the ASKAP image of each epoch with respect to their SUMSS counterpart. Red points and crosses represents the mean and standard deviation of the offsets which are also summarized in Table \ref{table:tests}. These offsets are fractions of the synthesized beam ($17\arcsec\times$ $12\arcsec$) and thus do not affect our analysis.}
\label{figure:imageoffsets}
\end{figure*}

\section{Observations and Data Reduction}\label{data reduction}
ASKAP is a radio interferometer currently being commissioned at the Murchison Radio-astronomy Observatory \citep{Johnston2007}. The Early Science program involves observations with a sub-array of twelve dishes of 12~m diameter equipped with Mk II phased array feeds (PAFs) \citep{hampson2012askap}.

A series of observations of a field centred on the interacting galaxy group NGC~7232 at RA: 22:08:03 and DEC: $-$44:22:55 (J2000) were conducted in 2016 October using 12 ASKAP dishes, with 36 PAF beams covering 30~deg$^{2}$ on the sky. The observations were performed at a central frequency of 1.4~GHz using a bandwidth of 48~MHz, divided into 2592 channels with a frequency resolution of about 18.519~kHz. Visibilities were recorded every 5 seconds.

The 36 PAF beams were aligned in a square 6$\times$6 footprint with pair of interleaving\footnote{\url{http://www.atnf.csiro.au/projects/askap/memo015.pdf}} configurations `A' and `B'. These interleaving configurations are a sequence of pointings with beam centres in the sensitivity depression of the previous pointing to achieve more uniform noise across the region.

The field was observed for 135 hours, divided into 12 epochs with approximately daily cadence yielding a typical rms noise of $\sim 300~\upmu$Jy/beam. In addition to the target field, the calibrator PKS~B1934$-$638 was observed with the same footprint pattern. The calibrator scan was conducted so that the calibrator source was observed at the centre of each of 36 PAF beams for 5 minutes. This scan was used for the bandpass calibration of each beam and also for setting the flux density scale.

We have used 8 epochs observed in the pointing ``B'' of the interleaving configuration, as our analysis identified flux scale discrepancies between sources observed in a mix of two configurations, which resulted in spurious variability. This could be because of the uncertainties in our knowledge of the primary-beam correction, which may lead to inconsistencies in the fluxes from the two footprints. Table~\ref{table:Measurement1} presents the details of these 8 observations. We note that the data products used in this analysis were obtained and processed in late 2016/early 2017 and are therefore different from the data products that are publicly available on the CASDA\footnote{\url{https://data.csiro.au/dap/public/casda/casdaSearch.zul}} data archive. 

\subsection{Data reduction} 
The visibility data for the calibrator and target field were written to measurement sets and then transferred to the Pawsey Supercomputing Centre, a national high-performance computing facility located in Perth, Western Australia as part of the ingest pipeline. The data were reduced using the ASKAPsoft pipeline version 0.17.0 \footnote{\url{ http://doi.org/10.4225/08/58e19e2904a45}}

\subsection{Splitting, flagging and bandpass calibration}
The calibrator and science field measurement sets were split by beam into 36 datasets using the ASKAPsoft task {\sc{mssplit}}. The calibrator and science data were then flagged using the task {\sc{cflag}}, which performed dynamic amplitude flagging. Autocorrelation products, bad antennas, radio frequency interference (RFI) affected channels and baselines were also flagged in this process.

The ASKAPsoft task {\sc{cbpcalibrator}} was used to perform the bandpass calibration. This calibration also sets the flux density scale to that of the Reynolds (1994)\footnote{\url{http://www.atnf.csiro.au/observers/memos/d96783~1.pdf}} model of PKS~B1934$-$638. The output solutions were written to a bandpass table, which were then applied to the science field using the task {\sc{ccalapply}}. The calibrated science target visibilities were then averaged in frequency to obtain measurement sets with 1~MHz channel widths. Another round of dynamic flagging was performed on the averaged measurement set and the science target visibilities were imaged.

\subsection{Continuum imaging and Self-calibration}
The 36-beam measurement sets were imaged separately using the task {\sc{cimager}} at their respective beam centres, derived from the footprint. {\sc{Cimager}} was used to perform imaging and deconvolution operations. Next, the process of self-calibration was performed using task {\sc{ccalibrator}}.
The imaging process is summarized as:
\begin{enumerate}
\item The bandpass-calibrated science target measurement sets for each beam were imaged in a parallel process;
\item The ASKAPsoft source finding algorithm {\sc{selavy}} \citep{Selavy} was used to produce a catalogue of sources with S/N $\geq 5$;
\item The resulting catalogue of sources was then used to create a model;
\item The model was used for refining gain solutions in the self-calibration step;
\item The refined gain solutions were applied during the next imaging iteration of the UV data;
\item Steps (ii) to (v) were repeated for two self-calibration loops.
\end{enumerate} 
Finally, after imaging each beam, the individual images were mosaicked together using the task~{\sc{linmos}}. The primary beam correction is performed at this stage, assuming a circular Gaussian beam. This may lead to small systematic errors in the flux densities of the sources, particularly at larger distances from the beam centres. Figure \ref{figure:image} shows the median image with an rms noise of $\sim 170~\upmu$Jy/beam, obtained by stacking seven images of similar shape and interleaving configuration.

\begin{table*}
\centering
\caption{Results of the astrometric and flux density calibration tests. Column 1 and 2 shows the epoch number and the number of ASKAP to SUMSS cross-matched sources used in these tests. Column 3 and 4 lists the mean positional offsets in right ascension and declination along with their standard errors. Column 5 presents the mean peak flux density ratio and the standard error. The flux densities have been corrected for the difference in frequencies assuming a spectral index of $-0.7$ as discussed in Sec \ref{photometry}.}
\label{table:tests}
\begin{tabular}{|c|c|cc|c|}
\hline
\multirow{2}{*}{Epoch no.} & \multirow{2}{*}{No. of sources} & \multicolumn{2}{c|}{Positional offsets} & \multirow{2}{*}{$S_{\rm ASKAP}/S_{\rm SUMSS}$ }  \\
& & RA & DEC  &  \\
\hline
\hline
1 & 360 & \phantom{---}0.4$\arcsec\pm 0.1\arcsec$ & \phantom{---}3.0$\arcsec\pm 0.1\arcsec$   & $0.97\pm 0.01$\\
2 & 340 & \phantom{---}0.2$\arcsec\pm 0.1\arcsec$ & \phantom{---}2.4$\arcsec \pm 0.1\arcsec$  & $0.97\pm 0.02$\\
3 & 281 & \phantom{---}2.2$\arcsec\pm 0.1\arcsec$& \phantom{---}3.0$\arcsec \pm 0.1\arcsec$   & $0.94\pm 0.02$\\
4 & 269 & \phantom{---}0.6$\arcsec\pm 0.1\arcsec$ & \phantom{---}2.0$\arcsec \pm 0.1\arcsec$  & $0.92\pm 0.02$\\
5 & 320 & \phantom{-}$-$0.3$\arcsec\pm 0.1\arcsec$ & \phantom{---}3.6$\arcsec\pm 0.1\arcsec$  & $0.95\pm 0.02$\\
6 & 348 & \phantom{-}$-$1.8$\arcsec \pm 0.1\arcsec$ & \phantom{---}4.1$\arcsec \pm 0.1\arcsec$& $1.00\pm 0.02$\\
7 & 216 & \phantom{---}3.4$\arcsec\pm 0.1\arcsec$ & \phantom{---}4.1$\arcsec \pm 0.1\arcsec$  & $0.92\pm 0.02$\\
8 & 241 & \phantom{---}2.0$\arcsec \pm 0.1\arcsec$ & \phantom{---}3.8$\arcsec \pm 0.1\arcsec$ & $0.95\pm 0.02$\\
\hline
\hline
\end{tabular} 
\end{table*}


\section{Image Analysis}\label{analysis}
We investigated the astrometric and flux density calibrations of our images by comparing the properties of sources detected in individual ASKAP epoch image with their Sydney University Molonglo Sky Survey (SUMSS) source catalogue counterparts \citep{sumss}. The SUMSS survey covered the sky south of declination $\delta = -30^{\circ}$ at 843~MHz with a resolution of 45$\arcsec\times45\arcsec \rm cosec\delta$ and a 5$\sigma$ flux density limit of 7.5~mJy at $\delta =-45^{\circ}$.
\subsection{Astrometry} 
\begin{figure*}
\includegraphics[scale=0.4]{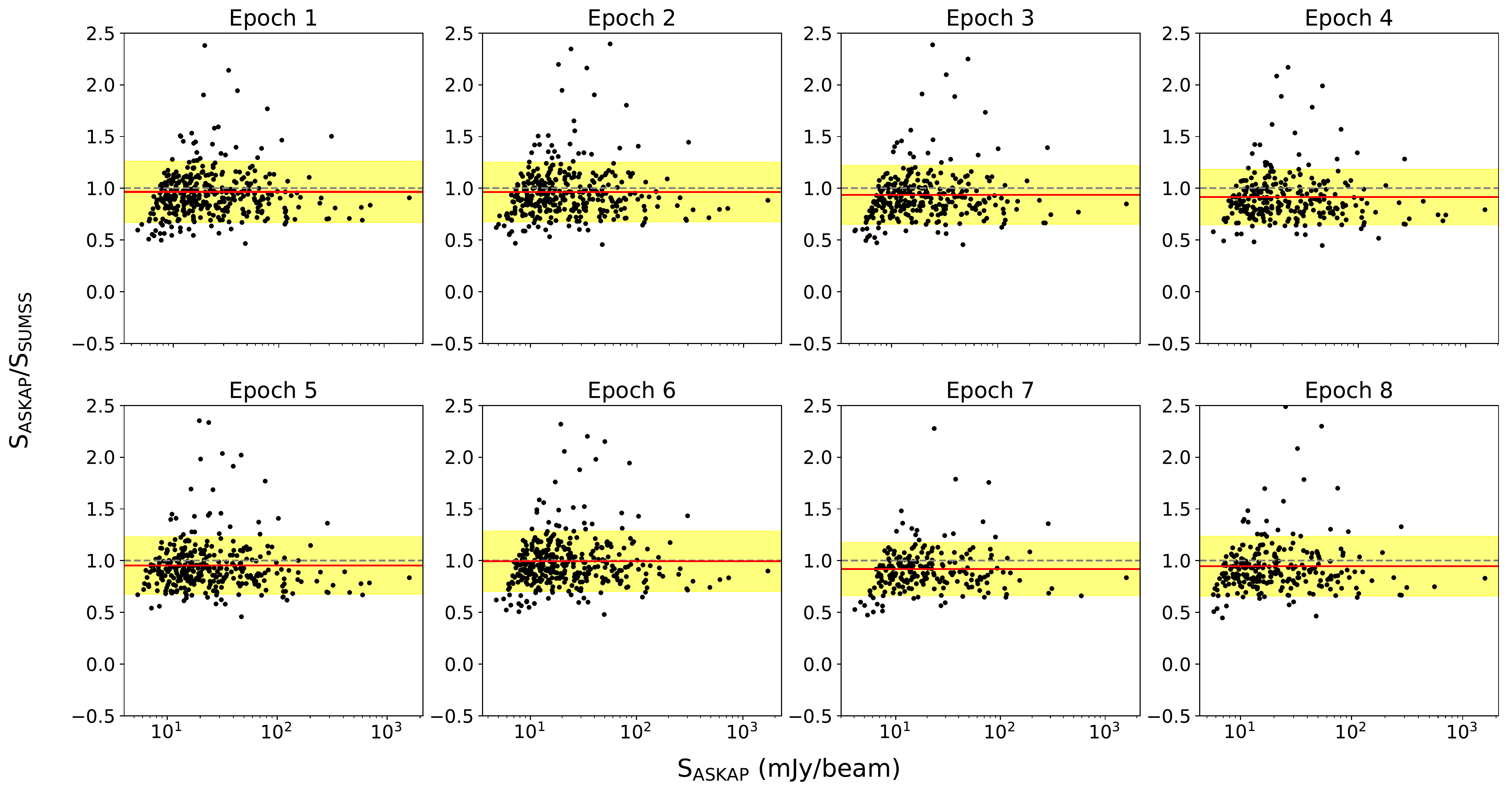}
\caption{The ratio of ASKAP to SUMSS peak flux density for compact and bright sources detected in ASKAP images. The mean of the ratio is listed in Table \ref{table:tests} and shown as a solid red line. The shaded region shows the standard deviation around the mean value. The flux ratio is consistent with 1.0 to within 10$\%$ between epochs. }
\label{figure:fluxstable}
\end{figure*}
The sources detected in ASKAP images with signal-to-noise ratio (S/N) $>$ 20 (bright) and a ratio of integrated flux density to peak flux density $\leq$ 1.3 (compact) were cross-matched with the sources in SUMSS catalogue. A search radius similar to the major axis of the synthesized ASKAP beam was used to perform cross-matching.  

The mean and standard deviation of offsets in right ascension (RA) and declination (DEC) for these compact and bright sources are summarized in Table \ref{table:tests} and shown in Figure \ref{figure:imageoffsets}. The offsets range from $-$1.8$\arcsec$ to 3.4$\arcsec$ in RA and 2.0$\arcsec$ to 4.1$\arcsec$ in DEC. These offsets are now understood as systematic errors that have been fixed. However, these positional offsets are fractions of the ASKAP synthesized beam width ($17\arcsec \times 12\arcsec$) and less than a pixel size of the ASKAP image ($4\arcsec$), therefore do not affect our variability analysis.
\begin{figure}
\includegraphics[scale=0.5]{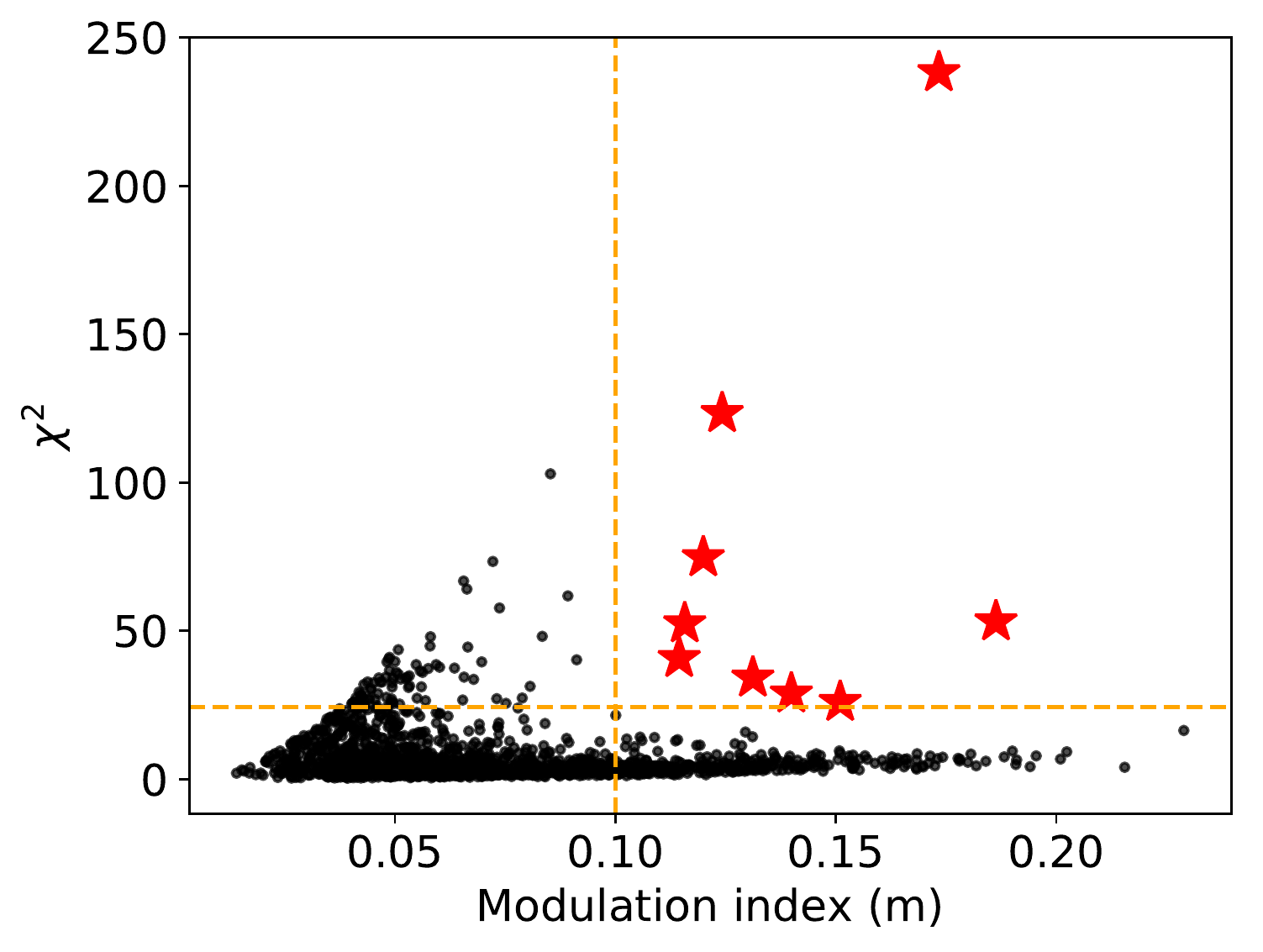}
\caption{The relationship between $\chi^{2}$ and $m$ is presented for 1653 sources detected in ASKAP epochs. The orange dashed line represent the $\chi_{T}^{2} = 24.3$ for 8 degrees of freedom and the modulation index cutoff $m=0.1$, the criteria for variable sources. Red stars are the potential variable sources detected in our analysis.} 
\label{figure:stats}
\end{figure}
\begin{figure*}
\includegraphics[scale=0.4]{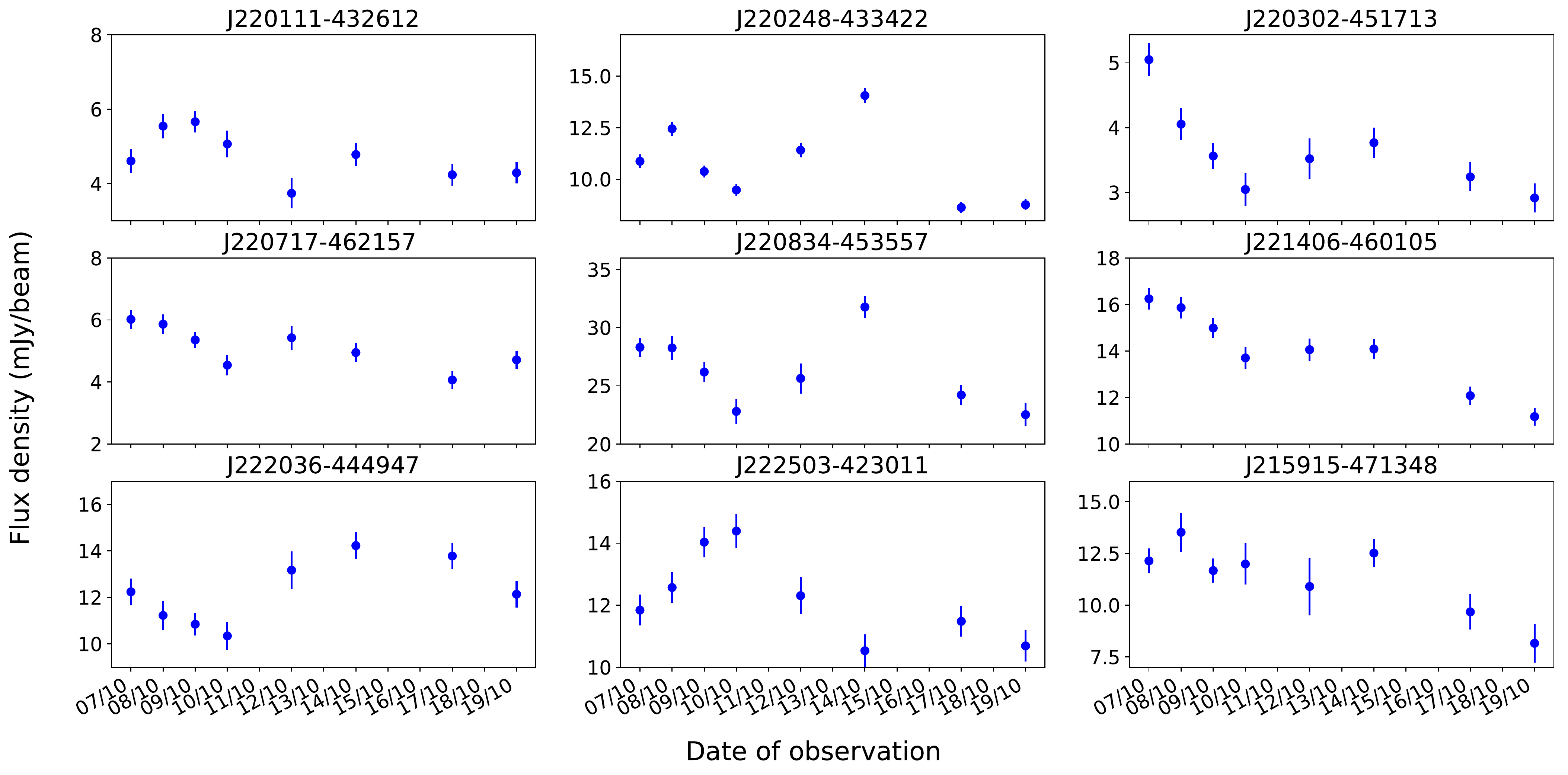}
\caption{The light curves of the 9 potential variable sources detected by the VAST pipeline in ASKAP observations at 1.4~GHz. The observed variability is consistent with the refractive scintillations of an active galactic nuclei.} 
\label{light}
\end{figure*}
\begin{figure}
\includegraphics[scale=0.27]{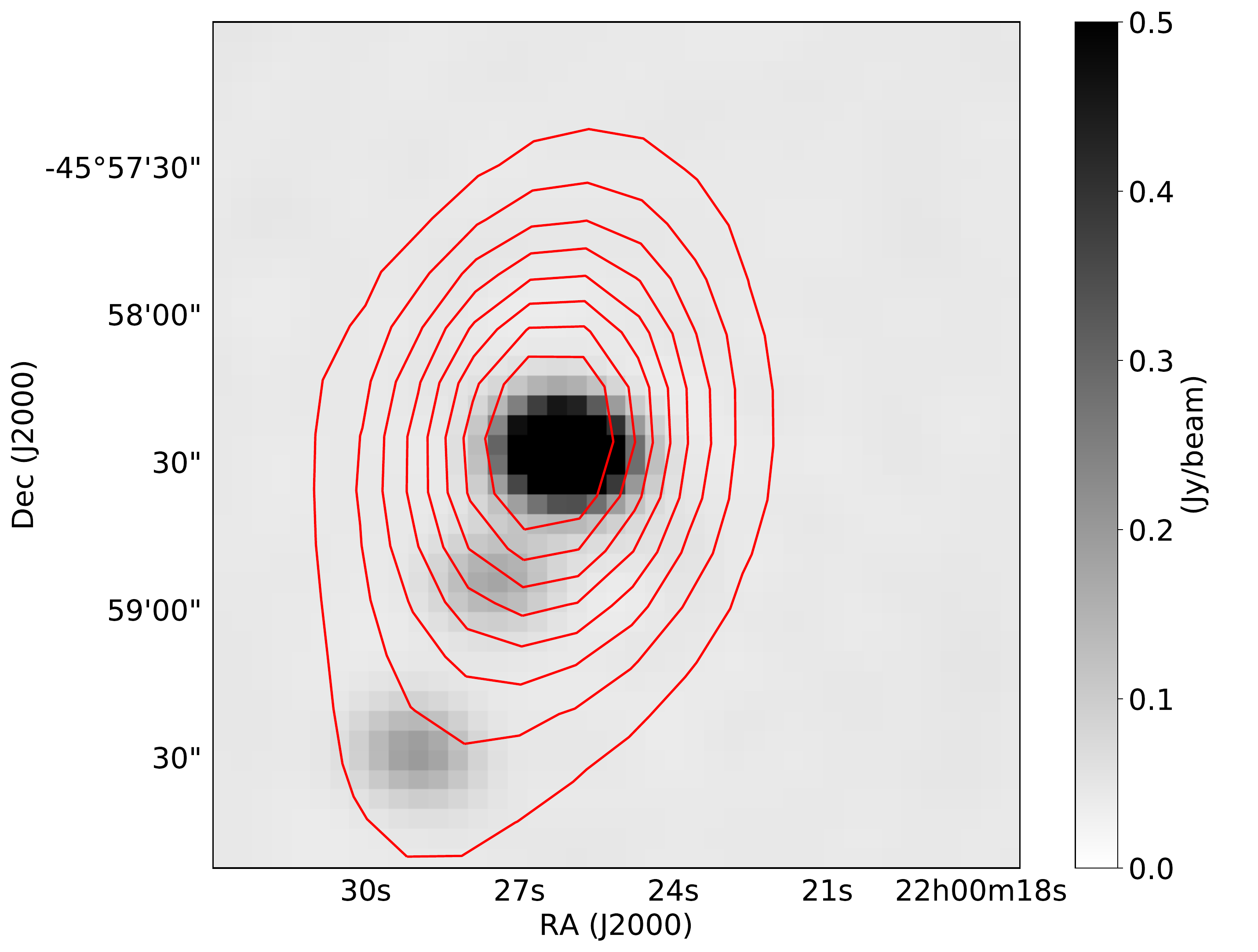}
\caption{A multi-component source detected in the ASKAP image, shown by the greyscale image. Over-plotted are the radio contours from SUMSS at 843 MHz with contour levels: 0.05, 0.10, 0.15, 0.20, 0.25, 0.30, 0.35, 0.40 in Jy. The source is resolved in the ASKAP image and hence we exclude such sources as potential transient candidates.} 
\label{multicomponent}
\end{figure}
\begin{figure*}
\centering
\subfigure[]{%
\includegraphics[scale=0.4]{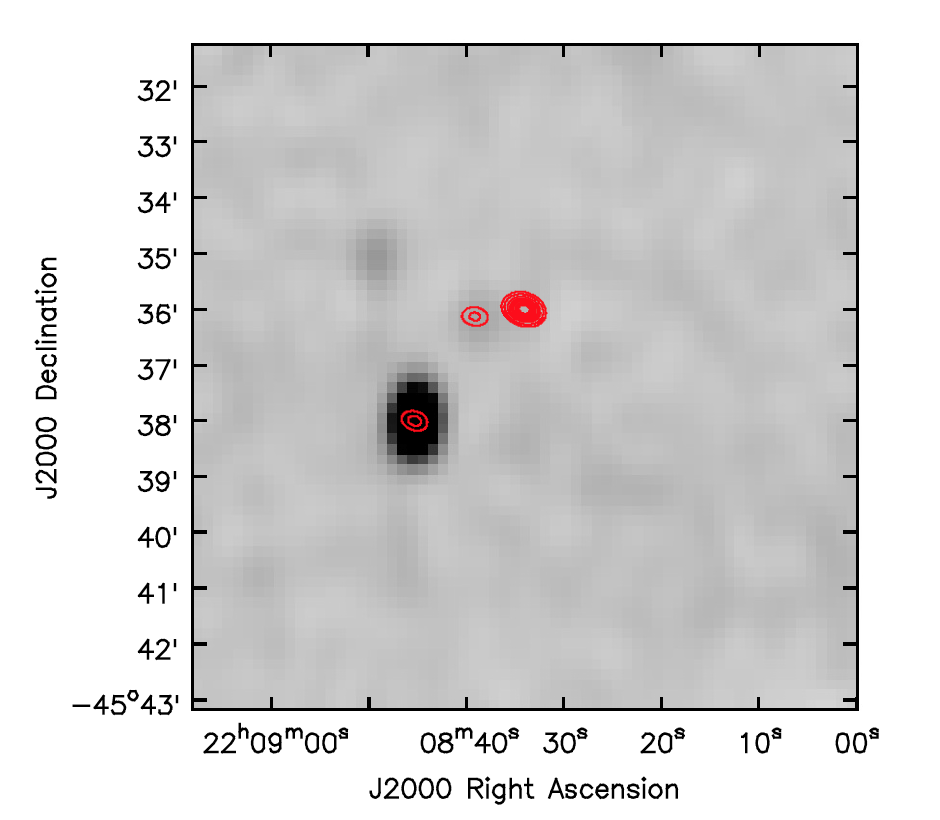} 
\label{summs_atca}}
\quad
\subfigure[]{%
\includegraphics[scale=0.4]{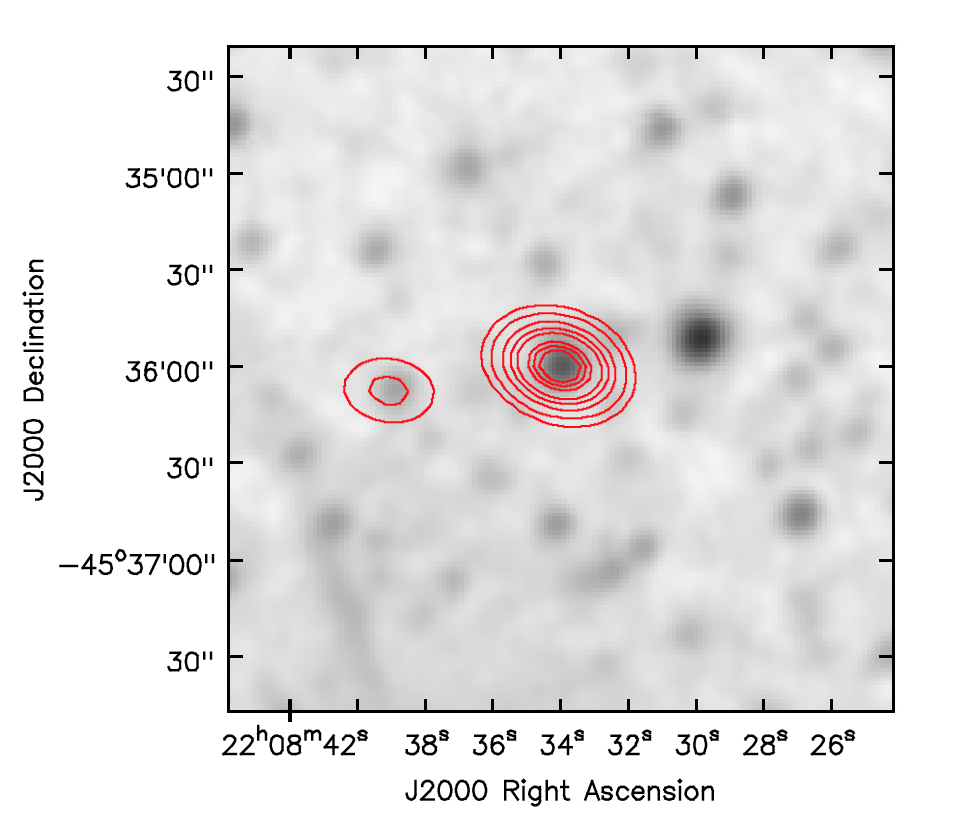} 
\label{figure:WISE_overlay}}
\caption{ Left panel: The SUMSS greyscale image overlaid with ATCA radio contours at 9~GHz in levels of 1.3~mJy, 2.7~mJy, 5.4~mJy, 8.2~mJy, 10.9~mJy, 13.6~mJy, 16.4~mJy, 19.1~mJy and 21.8~mJy. Right panel: The WISE greyscale zoomed-in image of the source J220833$-$453600 overlaid with ATCA radio contours at 9~GHz with the same contour levels as above.}
\label{figure:transientoverlays}
\end{figure*}

\subsection{Flux density calibration}\label{photometry}
We examined the flux density stability of the system by testing the absolute and relative flux density calibration. The bright compact sources detected in each ASKAP epoch (S/N $>$ 20 and a ratio of integrated flux density to peak flux density $\leq$ 1.3) were cross-matched to the SUMSS catalogue. We assumed a mean spectral index correction $\alpha=-0.7$ (S $\propto \nu^{\alpha}$, where $\alpha$ is the spectral index, $\nu$ is the observing frequency and $S$ is the flux density) to compensate for the frequency offset between the flux density of SUMSS and ASKAP sources at 843~MHz and 1.4~GHz respectively. The peak flux density ratios of cross-matched sources for each ASKAP epoch are presented in Table \ref{table:tests} and shown in Figure \ref{figure:fluxstable}. The absolute and relative (epoch to epoch) calibration of the system is within the 10\% level and we compensate for this error in our variability analysis.

\section{Results}\label{results}
\subsection{Variability Search}
\noindent The search for variables and transients was performed using the automated Variable and Slow Transient (VAST) pipeline \citep{Banyer2012}. The source finding algorithm \textit{Aegean} \citep{Hancock2018} was used for source detection and flux estimation. The software {\sc{bane}}\footnote{\url{https://github.com/PaulHancock/Aegean/wiki/BANE}} was used for estimating the background and rms images. We found 1653 sources common to 8 epochs and above a 5$\sigma$ flux density limit of $\sim$1.5~mJy, where $\sigma$ is rms noise of the image, and conducted a variability search on these sources. We used the chi-square ($\chi^{2}$) statistic to identify variable sources while the modulation index ($m$) 
was used to measure the degree of variability \citep{bell,Sadler2006}. They are defined as  
\begin{equation}
\chi^{2} =  \sum_{i=1}^{n} \frac{(S_{i}-\overline{S}_{\rm wt})^{2}}{\sigma_{i^{2}}}, 
\end{equation}

\begin{equation}
m = \frac{{\sigma_{S}}} {{\overline{S}}},
\end{equation}
where $S_{i}$ is the flux density in epoch $i$, $\overline{S}$ is the mean flux density, $\sigma_{i}$ is the uncertainty in $S_{i}$ and
$\overline{S}_{\rm wt}$ is the weighted mean flux density given by
\begin{equation}
\overline{S}_{\rm wt} = \sum_{i=1}^{n} \Big(\frac{S_{i}}{\sigma^{2}_{i}}\Big)  /\sum_{i=1}^{n} \Big(\frac{1}{\sigma^{2}_{i}}\Big).
\end{equation}

In the absence of variability, the values of $\chi^{2}$ are expected to follow the theoretical distribution $\chi^{2}_{\rm T}$ for $n-1$ degrees of freedom. We have calculated the cumulative distribution function (CDF), the probability $P$ of obtaining a $\chi^{2}$ value by chance. We consider a source to be variable if the $\chi^{2} > \chi^{2}_{\rm T}$ for P $<$ 0.001 (99.9\% confidence level). We initially found an excess of sources with large $\chi^{2}$ values, implying our flux density errors were underestimated. We therefore introduced an error of 2\% in the peak flux in quadrature with rms and source fitting errors to shift the mean of the reduced $\chi^{2}$ distribution to 1.0. After this correction, we recovered 52 candidate variable sources with modulation index $m > 0.1$. Sources with lower modulation indices are not reliably identified as variables. This also allows for the 10\% discrepancies in the flux calibration of the system.

After manual inspection of the 52 candidates, 9 sources were found to be potential variables as shown in Fig \ref{figure:stats}. The other sources were rejected due to the following reasons:
\begin{enumerate}
\item The source was extended; 
\item The source contained multiple components;
\item The source was detected at the edges of the image;
\item The source was coincident with imaging artefacts.
\end{enumerate}
The light curves of our 9 potential variable sources are presented in Fig \ref{light}. We analyze each of these sources in Section \ref{follow-up}. The variability statistics are listed in Table \ref{table:stats}.

We investigated beam-related systematics by checking the variable sources against their position with respect to the nearest beam. Variable sources J220716$-$462159 and J220833$-$453600 which lie near the centre and the edge of the Full Width Half Maximum (FWHM) of a beam (Beam12), ie 0.24$^{\circ}$and 0.57$^{\circ}$ offset from the pointing centre respectively, were also detected to be variable in the individual beam image. We did not detect any pattern in the variability of sources near and far away from the beam centre within the FWHM of the beam. We restricted our analysis to the FWHM of the
respective beams as the uncertainty in flux measurements could arise because of the difference between the assumed beam (Gaussian) and the actual beam and this uncertainty is likely to be worse further away from the beam centre for each beam.

We also investigated the spatial correlations of all sources detected in each ASKAP image in terms of their modulation indices and ASKAP to SUMSS 
flux ratios. We did not detect beam specific systematics which could cause artificial variability in our analysis.

\begin{table*}
\centering
\caption{Properties of the potential variable sources detected by the VAST pipeline. Column 3,4 lists the coordinates of the sources. Columns 5 through 8 presents the values of the variability statistics used in this analysis. Column 9 lists the spectral index for these sources after fitting a power law to the ATCA observations over a 2~GHz bandwidth.} 
\label{table:stats}
\begin{tabular}{|c|c|c|c|c|c|c|c|}
\hline 
 Index & Name & RA & DEC  & $m$  & $\chi^{2}$ & reduced $\chi^{2}$ & $\alpha$  \\
 & & (J2000) & (J2000) &  & & & \\
\hline
\hline
S1 & J220110$-$432614  & 22:01:10.76$\pm$0.03 &$-$43:26:14.4$\pm$0.3$\arcsec$ &0.14  &28.81	&4.12  &\phantom{-}$-$0.63$\pm$0.11 \\

S2 & J220247$-$433424  & 22:02:47.74$\pm$0.01 &$-$43:34:24.5$\pm$0.1$\arcsec$ &0.17  &238.27 &34.04  &\phantom{---}NA\\

S3 & J220301$-$451714  & 22:03:01.73$\pm$0.03 &$-$45:17:14.7$\pm$0.3$\arcsec$ &0.19  &53.12  &7.59   &\phantom{---}NA\\

S4 & J220716$-$462159  & 22:07:16.77$\pm$0.02 &$-$46:21:59.9$\pm$0.2$\arcsec$ &0.13  &34.16	&4.88   &\phantom{-}$-$0.26$\pm$0.07\\

S5 & J220833$-$453600  & 22:08:33.91$\pm$0.01 &$-$45:36:00.4$\pm$0.1$\arcsec$ &0.12  &74.74	&10.68  &\phantom{---}NA\\

S6 & J221405$-$460109  & 22:14:05.40$\pm$0.01 &$-$46:01:09.5$\pm$0.1$\arcsec$ &0.12  &123.33 &17.62  &\phantom{-}$-$0.17$\pm$0.02\\

S7 & J222035$-$444949  & 22:20:35.99$\pm$0.02 &$-$44:49:49.3$\pm$0.2$\arcsec$ &0.11 	&40.76	&5.82  &\phantom{-}$-$0.39$\pm$0.05\\

S8 & J222502$-$423011  & 22:25:02.72$\pm$0.02 &$-$42:30:11.6$\pm$0.2$\arcsec$ &0.12 	&52.46	&7.49   &\phantom{---}0.40$\pm$0.03\\

S9 & J215914$-$471350  & 21:59:14.20$\pm$0.02
&$-$47:13:50.7$\pm$0.4$\arcsec$ &0.15  &25.98  &3.71 &\phantom{---}0.15$\pm$0.04\\

\hline
\end{tabular} 
\end{table*}

\subsection{Transient Search}
We performed a search for the transient sources with near-daily cadence in 8 ASKAP epochs and no single-epoch transients were detected on timescales of days. 

We also performed a search for transients on a timescale of 14 years by comparing sources detected above the flux density limit of 14~mJy in ASKAP images to their SUMSS catalogue counterparts. The ASKAP sources that are not detected in the SUMSS catalogue/image are considered transients on timescale of 14 years. This analysis resulted in 33 sources. After careful visual inspection of these sources, we found 31 to have multiple components (many-to-one match between ASKAP and SUMSS) and thus were rejected as false candidates; see Fig \ref{multicomponent} for an example.

One of the remaining two sources (J215553$-$460517) was nonetheless detected in the SUMSS image with a flux density of $\sim$7.8~mJy. The absence of this source in the catalogue may be explained by the flux density being close to the 5$\sigma$ detection threshold of SUMSS. The second source J220833$-$453600, which had an average flux density of $\sim$26~mJy in the ASKAP images, is detected at a 3$\sigma$ level of $3.5\pm 1.2$~mJy in SUMSS image (Fig \ref{summs_atca}), is also uncatalogued.
This source is also identified as a variable source on daily timescales by the VAST pipeline in the ASKAP variability analysis. We discuss the follow-up observations and interpretation of this source in \S\ref{2208} and \S\ref{transient}.

We conducted a search for sources present in the SUMSS catalogue, but not in ASKAP images above a 10$\sigma$ flux density threshold. We found 25 such sources. However, they were all either extended in nature, near the edges of the field or coincident with artefacts near bright sources in the ASKAP images. We reject these sources as candidates for transients in our analysis.

\subsection{Australia Telescope Compact Array (ATCA) follow-up} \label{follow-up}
We conducted follow-up observations of the potential variables with the  Australia Telescope Compact Array (ATCA) to study their compactness and spectral properties. The observations were performed using the ATCA 6A configuration at a centre frequency of 2.1~GHz with a bandwidth of 2~GHz. Source PKS~B1934$-$638 was used for bandpass and flux calibration, while source PKS~2213$-$456 was used for phase calibration. Each of the nine sources was observed for 40~mins. We performed the synthesis imaging using the standard data reduction steps in {\sc{miriad}} \citep{miriad}. The flux density of the source of interest was obtained using the task {\sc{imfit}}.

\begin{figure}
\includegraphics[scale=0.35]{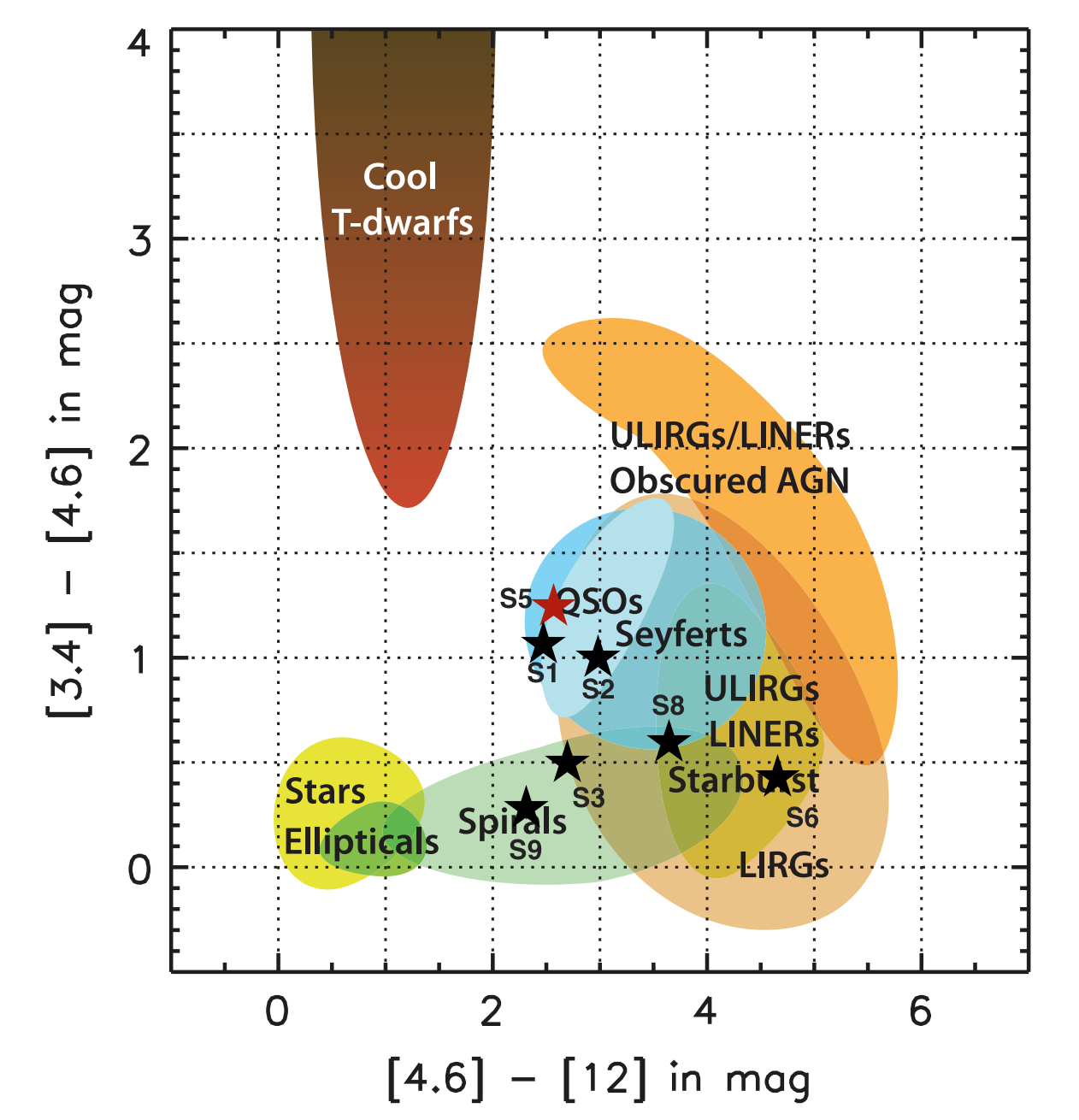}
\caption{The modified WISE colour plot from \citet{Wright2010}. The coloured regions indicate source classification. Black stars represent the potential variable sources detected in this analysis. The identifiers are used to map the sources to sources presented in Table \ref{table:stats}. The red star (S5) indicates the highly variable source J220833$-$453600.} 
\label{WISE_colour}
\end{figure}

The nine variable sources are compact in nature with the ratio of their integrated and peak flux density $\leq1.2$. Six of the nine sources are observed to have a flat spectral energy distribution (SED) with $-$0.3$~<\alpha<~$0.4 derived after fitting for a power law. These sources were queried in the Vizier\footnote{\url{http://vizier.u-strasbg.fr/viz-bin/VizieR}} archival data to search for their multi-wavelength counterparts. The search was performed using a radius of 5$\arcsec$. Most of the sources were cross-identified in the infra-red surveys such as the Wide-field Infrared Survey Explorer \citep[WISE;][]{Wright2010}, the Galaxy Evolution Explorer \citep[GALEX;][]{Bianchi2014} and the Two Micron All Sky Survey \citep[2MASS;][]{2MASS}. We searched the Guide Star Catalogue \citep[GSC;][]{Lasker1996}, USNO-B1.0 \citep{Monet2003} catalogues and the Sky-mapper optical database \citep{Wolf2018} to look for optical counterparts. Source J220716$-$462159 has no identified counterparts. The infra-red colours of the remaining sources are used and overlaid on a colour-colour plot in Fig~\ref{WISE_colour}. 

\subsection{Notes on individual sources}
\subsubsection{S1: J220110$-$432614} 
J220110$-$432614 matches WISE~J220111.10$-$432614.3 and the colours are consistent with a quasar. This source is also detected in 2MASS, GALEX, Million Optical-Radio/X-ray Associations Catalogue \citep[MORX;][]{MORX}, the DEep Near-Infrared Survey \citep[DENIS;][]{DENIS} and X-ray selected AGN in the 6dFGS-RASS catalogue \citep{RASS}. It is identified as a non-star in the GCS, has a Gaia counterpart \citep{Gaia2016} (ID 6570950867938602240) and is present in the Sky-mapper optical database (ID 307655951). It also matches an unresolved optical source in the USNO-B1.0 catalogue.

\subsubsection{S2: J220247$-$433424} 
J220247$-$433424 matches WISE~J220248.00$-$433424.1 and is consistent with the WISE colours of a quasar. There is no optical cross-match for this source. 
\subsubsection{S3: J220301$-$451714} 
J220301$-$451714 matches WISE~J220301.66$-$451715.1 and also detected in GALEX catalogue. Infra-red colours suggests this source is consistent with a spiral galaxy. It is classified as non-stellar source in GSC and also matches an unresolved optical source in the USNO-B1.0 catalogue.

\subsubsection{S5: J220833$-$453600} \label{2208}
\begin{figure*}
\centering
\subfigure[]{%
\includegraphics[scale=0.45]{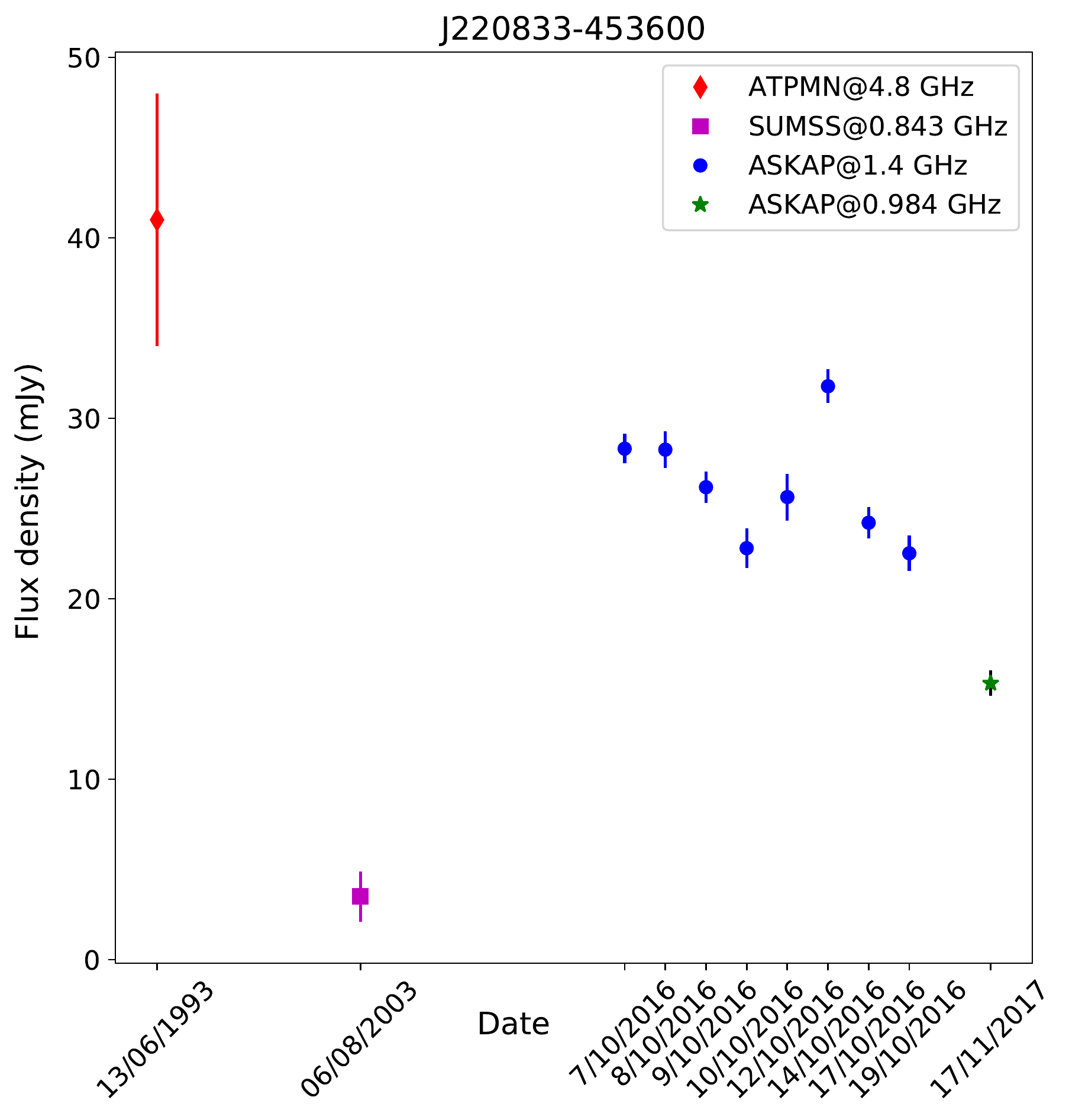} 
\label{figure:Transient_lightcurve}}
\quad
\subfigure[]{%
\includegraphics[scale=0.45]{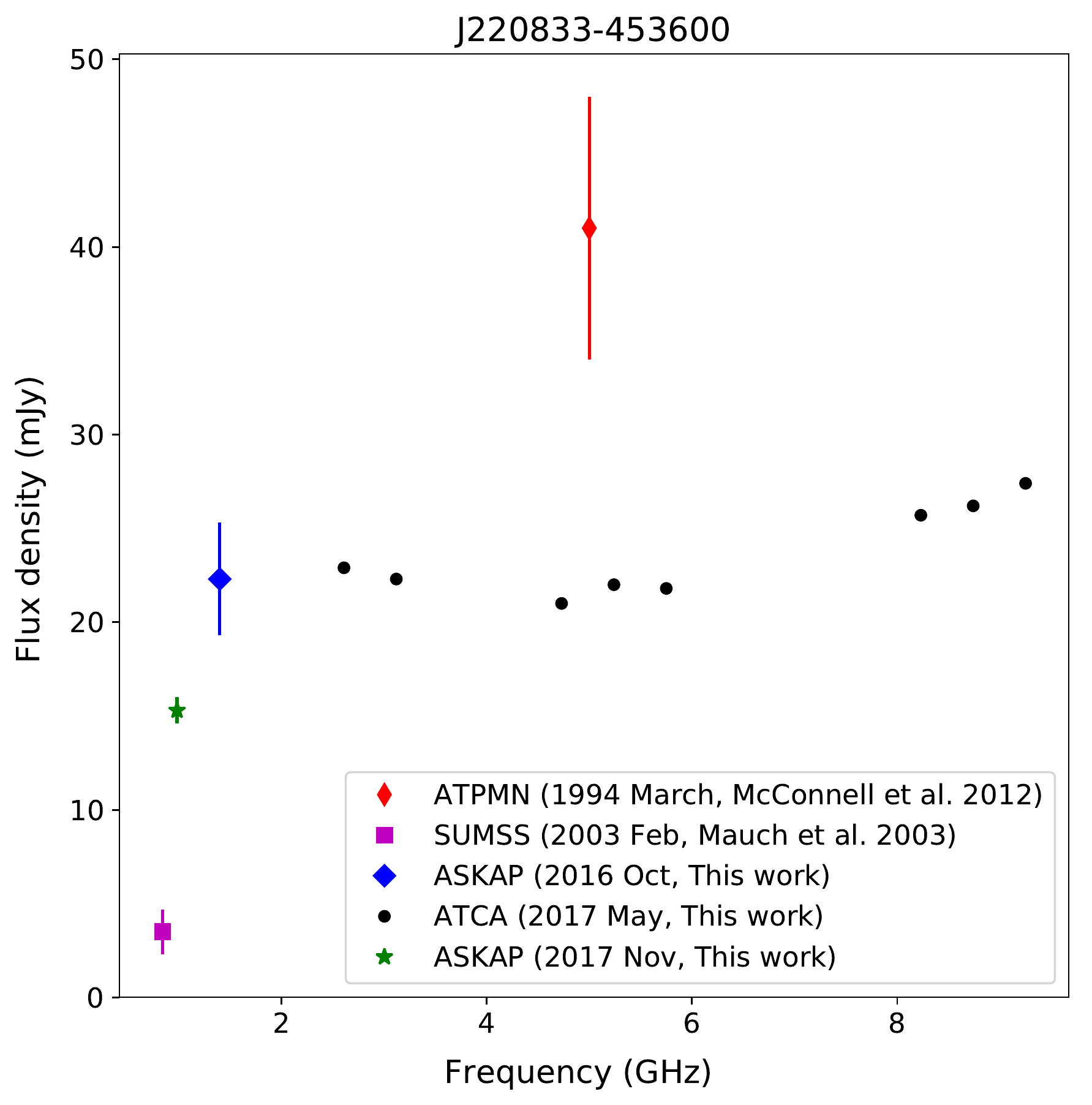} 
\label{figure:Transient_SED}}
\caption{ Left panel: The light-curve of the highly variable source J220833$-$453600. Right panel: The SED of the highly variable source J220833$-$453600. The errors on the flux densities are smaller than the points/symbols for ATCA observations in 2016. The ASKAP data point is the average flux density of 8 epochs and uncertainty is the scatter in flux density over 8 epochs.} 
\label{figure:transient}
\end{figure*}

J220833$-$453600 is a highly variable source with a modulation index $m=0.78$ between the SUMSS and ASKAP observations on the timescale of 14 years. It was also identified in the archival survey of the Australia Telescope Parkes-MIT-NRAO (ATPMN) at 5~GHz and 8~GHz \citep{ATPMN}. The source matches WISE~J220834.00-453559.4 (Fig. \ref{figure:WISE_overlay}) and infrared colours are consistent with a quasar. It is also detected in GALEX and AGN in the Mid-Infrared \citep{Secrest2015} catalogues, confirming the source to be an AGN. It is categorised as non-stellar in GSC, detected in the Sky-mapper optical database (ID 308634897) and also has a Gaia counterpart (ID 6567439130879039872).
\\
\\
\textbf{ATCA observations}: We observed the source J220833$-$453600 with the ATCA configuration H214 for 6~hrs with the centre frequencies of 2.1~GHz, 5.5~GHz and 9~GHz on 2017 June 25. The bandwidth of 2~GHz was further divided into 4 sub-bands. We excluded sub-band 1 which had a centre frequency of 1.9~GHz from our analysis as it was badly affected by RFI. Source PKS~B1934$-$638 was used for bandpass and flux calibration while source PKS~B2232$-$488 was used for phase calibration. We observed a $\sim$40$\%$ change in the flux density of the source at 5.5~GHz from 41~mJy in the ATPMN survey to 23~mJy in our latest ATCA observations. However, the flux density at 8~GHz is consistent with the ATPMN survey within the uncertainties. The spectrum of the source is observed to be flat at widely separated epochs of ATPMN observations (1993$-$1994) and our recent ATCA observations. 
\\
\\
\textbf{ASKAP observations}: We performed observations on 2017 Nov 11 with ASKAP at the centre frequency of 936~MHz and a bandwidth of 240~MHz. Source PKS~B1934$-$638 was used for bandpass calibration. The source was detected with a flux density of $\sim$15~mJy. The light curve of the source is presented in Fig~\ref{figure:Transient_lightcurve}. Since this source is also found to be variable on daily timescales, it is not certain if the flux measurement at 936~MHz suggests a spectral turnover as hinted in the spectrum in Fig \ref{figure:Transient_SED} or a variability effect. We discuss the possible causes of variability in Sec \ref{transient}.

\subsubsection{S6: J221405$-$460109}
J221405$-$460109 matches WISE~J221405.38$-$460107.1 and is consistent with the infra-red colours of a starburst galaxy. It is identified as non-stellar object in GSC. It also has a 
SUMSS counterpart with a flux density of 15.9$\pm$1~mJy and a 52\% probability of being a quasar and 17\% probability for a galaxy in the MORX catalogue. 
\subsubsection{S7: J222035$-$444949}
J222035$-$444949 is only identified in the AC 2000.2 catalogue \citep{Urban2001}. It does not have any infrared counterpart.
\subsubsection{S8: J222502$-$423011}
J222502$-$423011 matches WISE~J222502.65$-$423013.3 and is also consistent with the infra-red colours of a starburst galaxy. No optical counterpart was identified for this source.
\subsubsection{S9: J215914$-$471350}
J215914$-$471350 matches WISE~J215914.38$-$471350.8 and the colours are consistent with a spiral galaxy. This is also detected in 2MASS, DENIS, the Galaxy List for the Advanced Detector Era catalogue \citep[GLADE;][]{GLADE} and the 2dF Galaxy Redshift Survey \citep{Colless2003}, which classifies this source as a galaxy. This source is identified as non-stellar in GSC. 
We also searched the Sky-mapper data and found a match with an extended source (ID 308593832). 

We encourage optical observations of these sources for their true identification and multi-wavelength follow-up observations to monitor their variability. 

\begin{figure}
\includegraphics[scale=0.3]{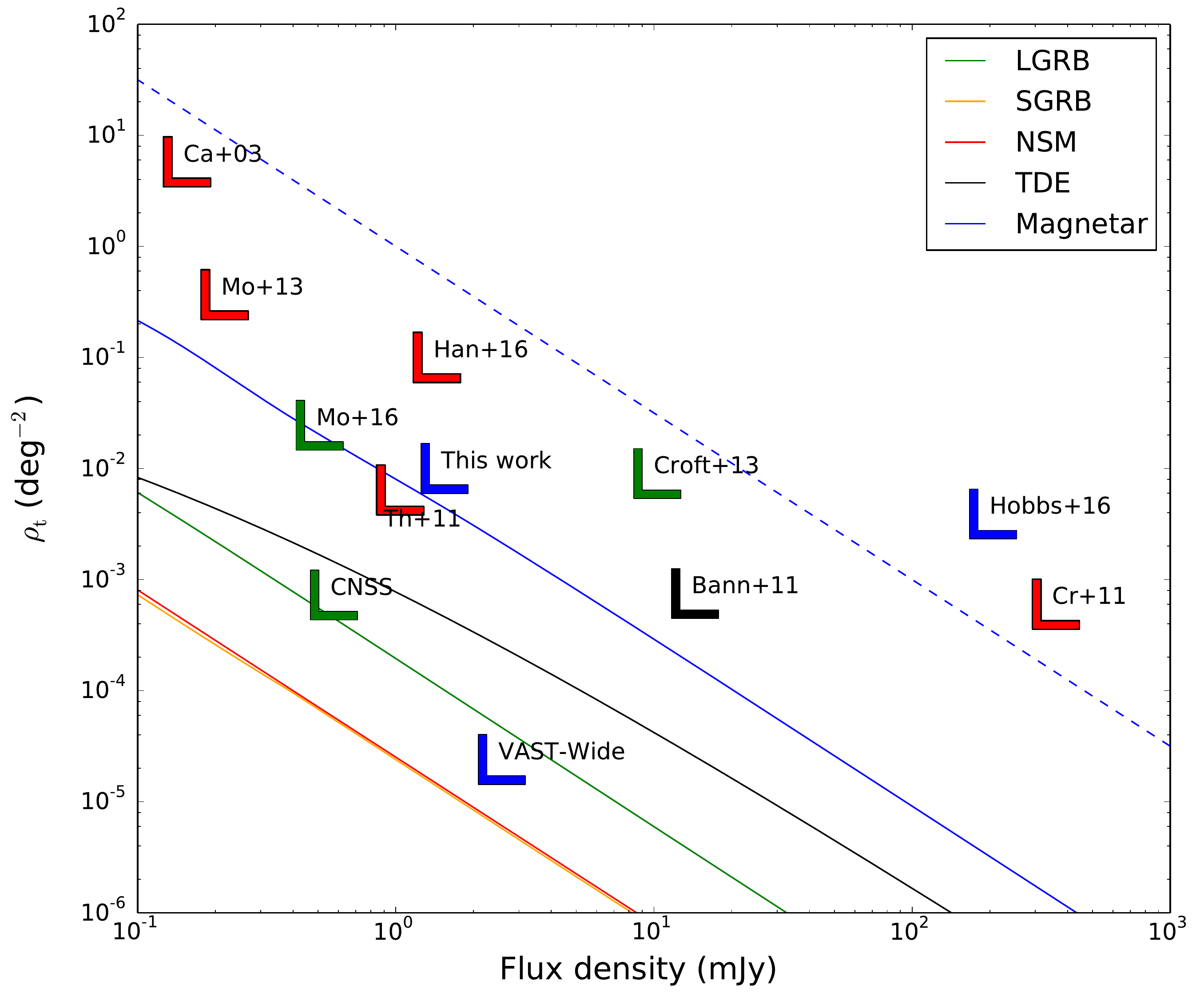}
 \caption{Transient source surface density limits for a range of current and planned surveys. Red symbols show surveys performed at 1.4~GHz (\citet{Carilli2003}, \citet{Mooley2013}, \citet{Thy2011}, \citet{Croft2011}, \citet{Hancock2016}), green symbols show the surveys performed at 3~GHz (\citet[][CNSS pilot, CNSS]{mooley} and \citet{Croft2013}). Blue symbols show the transient source densities obtained from ASKAP at 863.5~MHz by \citet{Hobbs2016}, at 1.4~GHz (This work) and predicted upper limits for VAST-Wide survey \citep{VAST}.
The black symbol is the survey performed by \citet{Bannister2011} at 843~MHz. We have used
the quoted flux density limits for respective surveys and they are plotted in
the x-axis.We have used $5\sigma$ flux density limit for the VAST-Wide survey for
comparison with our pilot survey. The dashed blue line shows $\rho_{\rm t} \propto S^{-3/2}$, the relation for a Euclidean source population.
Other coloured lines are the upper limits for models of neutron star mergers (NSM), magnetars, long and short gamma ray bursts (LGRB, SGRB) and tidal disruption events (TDE) \citep{Metzger}.
}
\label{figure:transient}
\end{figure}
\section{Discussion and Summary}\label{discussions}
\subsection{Variables on timescale of days}

Nine potential variable radio sources ($0.5\%$ of the total) were detected in our analysis, with flux density variations on the timescale of days. All sources showed low level variability with modulation indices $m$ ranging from 0.1 to 0.2. The brightness temperature can be used as a tool to test if the variability is intrinsic or extrinsic to the source. Assuming the source size and angle subtended by the source are limited by the light travel time, the brightness temperature and amplitude of the variability are related by:
\begin{equation}\label{brightness_temp}
T_{\rm B} \geq \frac{\Delta SD^{2}}{2k_{\rm B}\nu^{2}\tau^{2}},
\end{equation}
where $\Delta S$ is the amplitude of the variability, $k_{\rm B}$ is the Boltzmann constant, $D$ is the distance to the source, $\nu$ is the observing frequency and $\tau$ is the timescale of the variability. If the variability is driven by a synchrotron process, then a brightness temperature $T_{\rm CC}<10^{12}$~K satisfies the Compton catastrophe limit \citep{Kellermann1969}. Moreover, \citet{Readhead1994} showed that the brightness temperatures in powerful extragalactic radio sources is not always constrained by the Compton catastrophe limit and the ``equipartition brightness temperature" can be lower, $T_{\rm IC} < 3 \times 10^{11}$~K due to inverse Compton cooling. 
Using an average $\Delta S=4$~mJy in our sample, $\nu = 1.4$~GHz, assuming $D=10$~Gpc, we obtain a timescale 
$\tau=15.3$~yrs for $T_{\rm IC} = 3 \times 10^{11}$~K.
Since we observed the variability on the timescale of days in our sample, it is likely to be extrinsic. 

The variable sources detected in our analysis are identified to be quasars or galaxies hosting an AGN. Such compact sources could undergo refractive interstellar scintillation (RISS). The timescale associated with RISS scales with frequency as $\nu^{-\frac{\beta}{\beta-2}}$, where $\beta=11/3$ for Kolmogorov turbulence \citep{Rickett1977}. Thus, the timescale of RISS for a source which shows intra-day variability at 2~GHz is $\sim$2~days at 1.4~GHz. RISS of an AGN could therefore explain the observed variability in our sample, which is also consistent with the previous studies \citep{Gaensler2000,Rickett2006,Ofek2011}. We also estimated the expected modulation for RISS using the equations 11-13 in \citet{Walker1998}. We obtain a modulation index of $\sim$0.4 with the timescale of $\sim$4~days using a transition frequency of 8~GHz at a Galactic latitude of $-50^{\circ}$. This is consistent with the modulation indices obtained for variable sources in our pilot survey. 

\subsection{Variable on timescale of a decade}\label{transient}
By comparing the sources detected in the ASKAP images with their SUMSS counterparts, we have probed variability on timescales of a decade. J220833$-$453600 was discovered to be a highly variable source in this analysis. Possible reasons for the cause of variability in the source J220833$-$453600 are
\begin{itemize}
\item long-term intrinsic variability of an AGN,
\item an explosive flare,
\item an extreme scattering event (ESE),
\item variability due to refractive scintillation.
\end{itemize}
The fact that the SUMSS epoch date is between the ATPMN and ASKAP dates suggests long-term variability rather than a single flare or explosive transient. Certainly the relatively high flux density and lack of evolution in the radio spectrum is inconsistent with a radio supernova or GRB afterglow. The covering fraction of ESEs is $\sim 1/2000$ sources at a given time and they are stronger at 5~GHz. Therefore, the probability of an ESE nulling out the SUMSS observation is relatively low.
The long-term intrinsic variability of an AGN, particularly that of a quasar, seems to be the most likely explanation for the observed variability. This is backed up by the timescale calculation obtained using Eq.\ref{brightness_temp}, which is consistent with the observed timescale of the long-term intrinsic variability of an AGN. This source is also observed to be variable on daily timescales, suggesting RISS plays a role in the flux density variations. 

\begin{table*}
\centering
\caption{Comparison of the variable source densities for different surveys. Column 1 lists 
the surveys including this work and 5 others from the literature. Column 2 through 5 lists the flux density limits, frequencies, area surveyed and number of variable sources detected in these surveys. Column 6 and 7 list the variable source surface densities ($\rho_{\rm v}$), with 1$\sigma$ Poisson error calculated following \citet{Gehrels}, and finally, the sampling timescales for each survey.}
\label{table:variable density}
\begin{tabular}{|c|c|c|c|c|c|c|c|c|}
\hline 
Survey & Flux limit& Frequency  & Area & No. of variables & $\rho_{\rm v}$& timescale \\
& (mJy) & (GHz) &  (deg$^{2}$)& sources & (deg$^{-2}$) & \\
\hline
\hline
This work & 1.5 & 1.4 & 30 & 9 & 0.3$^{+0.1}_{-0.1}$ & days \\ 
This work & 1.5 & 1.4 & 30 & 1 & 0.03$^{+0.07}_{-0.02}$ & decade \\ 
\citet{mooley} & 0.5 & 3 &50 & 38&0.76$^{+0.14}_{-0.12}$ & weeks \\
\citet{mooley}  & 0.5 & 3& 50&31 &0.62$^{+0.13}_{-0.11}$ & months \\
\citet{bell} & 0.086 & 5.5 & $\sim$0.3 &4 & 13.3$^{+10.4}_{-6.3}$ & month to years \\
\citet{Hancock2016} & 1.4 & 1.4 & $\sim$8 & 8 & 0.98$^{+0.5}_{-0.34}$ & 6-month to years \\
\citet{mooley} &  0.5 & 3& 50& 96&1.92$^{+0.2}_{-0.2}$ & years \\

\hline
\end{tabular} 
\end{table*}

\subsection{Comparison with previous surveys}
The variable source surface densities for the current survey are $\rho_{\rm v}=0.3^{+0.1}_{-0.1}$~deg$^{-2}$ and $\rho_{\rm v}=0.03^{+0.07}_{-0.02}$~deg$^{-2}$ on timescales of days and decade respectively, above a flux density limit of 1.5~mJy. These are compared with the previous surveys of \citet{bell}, \citet{mooley} and \citet{Hancock2016} and are presented in Table \ref{table:variable density}.

The upper limit on the transient source surface density for no detections at the 95\% confidence limit is given by 
\begin{equation}
\rho_{\rm t} < \frac{-ln(0.05)}{(n-1) \times \Omega}, 
\end{equation}
where $\rho_{\rm t}$ is the transient source surface density, $n$ is the number of epochs and $\Omega$ is the sky area surveyed in deg$^{2}$. The transient source surface density for our pilot survey covering 30~deg$^{2}$ of the sky, with no transient detected on daily timescales is $\rho_{\rm t} < 0.01$ deg$^{-2}$ at 95$\%$ confidence. We compare the surface density for a range of transients with \citet{mooley} in Table \ref{table:mooley_comparison}. We assume a slope of the source count distribution (logN-logS) of the extragalactic sources as $\sim-$2.0 \citep{Condon1988} in our calculations. The limits for the range of transients are not well constrained. However, we did expect to detect an AGN undergoing interstellar scintillation in our survey, which is consistent with our findings.

Fig. \ref{figure:transient} shows the predicted upper limits/rates of transient source surface densities for various physical processes calculated by \citet{Metzger}, including neutron star merger (NSM), magnetars, tidal disruption events (TDEs), short and long gamma ray bursts (GRBs). The non-detection of the afterglows related to the physical mechanisms mentioned above is consistent with the predictions of \citet{Metzger}. 

We also estimate the transient source surface density for the VAST-Wide survey \citep{VAST} as $\rho_{\rm t}~<~2.4 \times 10^{-5}$~deg$^{-2}$, at 95$\%$ confidence. This survey is planned to cover 10,000 deg$^{2}$ of sky with daily cadence for 2 years, with a sensitivity of 0.5~mJy/beam. The survey will detect rare and bright events such as GRBs, supernovae and monitor a large number of bright transients and variables such as intra-day variables (IDVs) and AGN.

\begin{table}
\caption{Modified version of the summary of slow radio transients in \citet{mooley}. Column 1 and 2 describe the object and their variability timescales. Column 3 is the source surface density scaled to the flux density limit of the present survey.}
\label{table:mooley_comparison}
\begin{tabular}{|c|c|c|}
\hline 
Object & Timescale  & Rate ($>$1.5~mJy) \\
& &   (deg$^{-2}$) \\
\hline
\hline
AGN (Shock-in-jet)  & days-years & 0.05 \\ 
AGN (ISS) & minutes-days & 2.4 \\
SN-Ia & days-week & $<4 \times 10^{-7}$ \\
Long-GRB  &days-years & $\phantom{-}2 \times 10^{-6}$\\
Short-GRB & days-years? &$<4 \times 10^{-8}$ \\
BNS merger & weeks-years &$\phantom{-}8 \times 10^{-6}$ \\
TDE  &years? & $\phantom{-}2 \times 10^{-4}$ \\
\hline
\end{tabular} 
\end{table}

\subsection{Future prospects}
The future VAST-Wide survey with a 36 dish ASKAP array, will be sensitive to exploring the parameter space predicted for number of transients such as GRBs, NSM, TDEs, magnetars, which are currently not probed by this pilot survey.
The wide field-of-view and $\sim 100~\upmu$Jy/beam sensitivity for a 5 minute observation, will also make ASKAP an excellent instrument for follow-up of events from the Laser Interferometer Gravitational-Wave Observatory (LIGO) \citep{LIGO}.
It will cover the typical LIGO localization error of 600-1600~deg$^{2}$ with 20-50 pointings to look for possible electromagnetic counterparts to gravitational waves. 
The upper limits for transients derived from this survey can be used for predicting the number of transients in the LIGO event localization region, such as $\leq6$ transients in 600~deg$^{2}$ area of sky for the GW~150914 event \citep{Abbott2016,Hotokezaka2016}.

\subsection{Summary}
We have presented a transient and variable source search with ASKAP over 8 epochs on a timescale of days covering 30~deg$^{2}$ of sky. Nine sources were found to display variability in their flux densities in observations separated by $\sim$24~hr. We reject the null hypothesis that the flux densities of these sources does not change, with 99.9\% confidence. We further discuss the plausible explanations of the variability and conclude that refractive interstellar scintillation of compact AGN is the most likely cause. No transients were detected on timescales of days. We also detected a highly variable source on a timescale of 14 years consistent with a long-term intrinsic variability of a quasar. We have shown that a shallow survey with ASKAP could discover a potential new class of highly variable sources when compared with SUMSS.  

We have demonstrated and tested the functionality of the VAST pipeline for ASKAP data. The transient upper limits obtained are already competitive with previous surveys, suggesting the final implementation of ASKAP will be probing a new phase space for transients.  

\section*{Acknowledgements}
SB would like to thank Igor Andreoni, Christian Wolf for useful discussions about optical counterparts and Brian Metzger for providing the data curves of predicted source surface densities for various transient processes. TM acknowledges the support of the Australian Research Council through grant FT150100099. DLK was supported by NSF grant AST-1412421. Part of this research was conducted by the Australian Research Council Centre of Excellence for All-sky Astrophysics (CAASTRO), through project number CE110001020. Parts of this research were conducted by the Australian Research Council Centre of Excellence for All-sky Astrophysics in 3D (ASTRO 3D) through project number CE170100013.

The Australian SKA Pathfinder is part of the Australia Telescope National
Facility which is funded by the Commonwealth of Australia for operation as a
National Facility managed by CSIRO. This scientific work uses data obtained
from the Murchison Radio-astronomy Observatory (MRO), which is jointly
funded by the Commonwealth Government of Australia and State Government
of Western Australia. The MRO is managed by the CSIRO, who also provide
operational support to ASKAP. We acknowledge the Wajarri Yamatji people as
the traditional owners of the Observatory site. The work was supported by
the Pawsey supercomputing centre through the use of advanced computing resources.

\bibliographystyle{mnras}
\bibliography{references}

\bsp	
\label{lastpage}
\end{document}